\documentclass[twocolumn,showpacs]{revtex4}
\usepackage[T1]{fontenc}
\usepackage{bm}
\usepackage{graphicx}
\usepackage{epstopdf}
\usepackage{amsmath}
\usepackage{siunitx}
\usepackage{color}

\begin{document}

\title{2$s$ exciton-polariton revealed in an external magnetic field}

\author{B.~Pi\k{e}tka$^1$}
\author{M.~R.~Molas$^2$}
\author{N.~Bobrovska$^3$}
\author{M.~Kr\'{o}l$^1$}
\author{R.~Mirek$^1$}
\author{K.~Lekenta$^1$}
\author{P.~St\k{e}pnicki$^3$}
\author{F.~Morier-Genoud$^4$}
\author{J.~Szczytko$^1$}
\author{B.~Deveaud$^{4,5}$}
\author{M.~Matuszewski$^3$}
\author{M.~Potemski$^{1,2}$}

\affiliation{$^1$Faculty of Physics, University of Warsaw, Warsaw, Poland}
\affiliation{$^2$Laboratoire National des Champs Magn\'etiques Intenses, CNRS-UGA-UPS-INSA-EMFL, Grenoble, France}
\affiliation{$^3$Institute of Physics, Polish Academy of Sciences, Warsaw, Poland}
\affiliation{$^4$Ecole Polytechnique F\'ed\'erale de Lausanne (EPFL), Lausanne, Switzerland}
\affiliation{$^5$Ecole Polytechnique, F-91128 Palaiseau, France}

\begin{abstract}
We demonstrate the existence of the excited state of an exciton-polariton in a semiconductor microcavity. The strong coupling of the quantum well heavy-hole exciton in an excited $2s$ state to the cavity photon is observed in non-zero magnetic field due to surprisingly fast increase of Rabi energy of the $2s$ exciton-polariton in magnetic field. This effect is explained by a strong modification of the wave-function of the relative electron-hole motion for the $2s$ exciton state.

\end{abstract}

\keywords{exciton-polaritons; semiconductor microcavity; photoluminescence; magnetic field}
\pacs{78.55.Cr, 71.35.Ji, 71.36.+c, 42.55.Sa, 73.21.Fg}


\maketitle
\section{\label{sec:Intro}Introduction}

The Coulomb bound electron-hole pairs, the excitons, exhibit an internal structure with a series of excited states below the continuum band edge. In a quantum well (QW), the energy structure of the exciton is similar to that of the two-dimensional hydrogen atom with the subsequent 1$s$; 2$s$, 2$p$; 3$s$, 3$p$, 3$d$ and higher excited states. The optically active states have an $s$ symmetry~\cite{Elliot} and the first demonstration of the strong coupling between the cavity photon and the 1$s$ exciton state was shown by Weisbuch et al. \cite{weisbuch}. Tignon et al. \cite{tignon} showed the appearance of the strong coupling of the 2D-continuum of excitations to the cavity photon upon increasing magnetic field. The transition from weak to strong coupling was observed by quantizing unbound excitonic states into Landau levels in the magnetic field.

Recent interest in higher excited state excitonic ladder in a microcavity~\cite{rupuhupu} motivated us to demonstrate the strong coupling of the 2$s$ excited state of the exciton to cavity photons. We observe a strong increase of the 2$s$ exciton - cavity photon coupling strength in magnetic field. The enhancement of the coupling strength is directly related to the increase of the exciton oscillator strength induced by a strong magnetic field. 

We find that the coupling of the 2$s$ exciton and the microcavity photon is hardly visible at zero magnetic field, but becomes increasingly pronounced at higher field strengths. The magnetic field action is twofold. First, it lifts the degeneracy between the excited 2$s$ and 2$p$ exciton states~\cite{a, b, c}. Second, it acts as an effective confining potential for the electron and the hole, increasing their wave function overlap. We note that the increase of the 1$s$ exciton ground state oscillator strength in magnetic field was already reported in~\cite{whittaker, fisher, berger, armitage, forchel, pietka}.

In this paper we demonstrate directly the strong coupling between the 2$s$ exciton state and a cavity photon by the examination of the dispersion of polariton photoluminescence (PL) in a wide range of emission angles and magnetic fields. We observe the fast increase of both the 2$s$ exciton energy and the exciton oscillator strength in magnetic field. The latter results in an enhancement of the Rabi energy, or the coupling strength between excitons and photons, at a much higher rate than in the case of the $1s$ exciton. This phenomenon is explained theoretically using a model of an artificial two-dimensional hydrogen atom. We find that the weakly bound $2s$ exciton is much more sensitive to magnetic field due to the modification of its relative electron-hole wave-function, which leads to a stronger diamagnetic effect. In the framework of the first-order perturbation theory, we explain the effect qualitatively by the proximity of higher excited states, which is especially pronounced in the two-dimensional case where the exciton is confined in a quantum well.

\section{Exciton-polariton system and experimental details}
\label{sec:exp}

Exciton-polaritons are the eigenmodes of a strongly coupled system composed of a photonic mode in a microcavity and an excitonic resonance inside a semiconductor QW~\cite{weisbuch}. Our sample is composed of a lambda GaAs microcavity sandwiched between two AlAs/GaAs distributed Bragg reflectors (DBR). One 8~nm-thick In$_{0.04}$Ga$_{0.96}$As QW was placed in the maximum of the cavity field providing the excitonic component to polaritons. Details on the sample structure can be found in Ref.~\cite{ounsi}. The exciton -- photon coupling strength is given by the Rabi energy, which is of 3.5~meV at zero magnetic field. The linewidth of the polariton modes at exact exciton -- photon resonance is 0.3~meV.  This narrow resonance linewidth, compared to the Rabi energy, allows us for a very clear resolution of the different polariton states.

The sample was placed in a superconducting magnet providing magnetic field up to 14~T, in a Faraday configuration, and at liquid helium temperature. The sample was excited non-resonantly by a cw Ti:Sapphire laser with the energy tuned to the first reflectivity minima of the DBRs on the high energy side. The photoluminescence was excited and collected through a high numerical aperture ($\textrm{NA}=0.83$) single lens of 9~mm focal length. The high angular resolution of our detection system allowed us,  within the same method as presented in Ref.~\cite{kasprzak}, to measure directly the dispersion of polaritons: the emission angle is proportional to the polariton in-plane momentum.

\begin{figure}
	\centering
	\includegraphics[width=8.7cm]{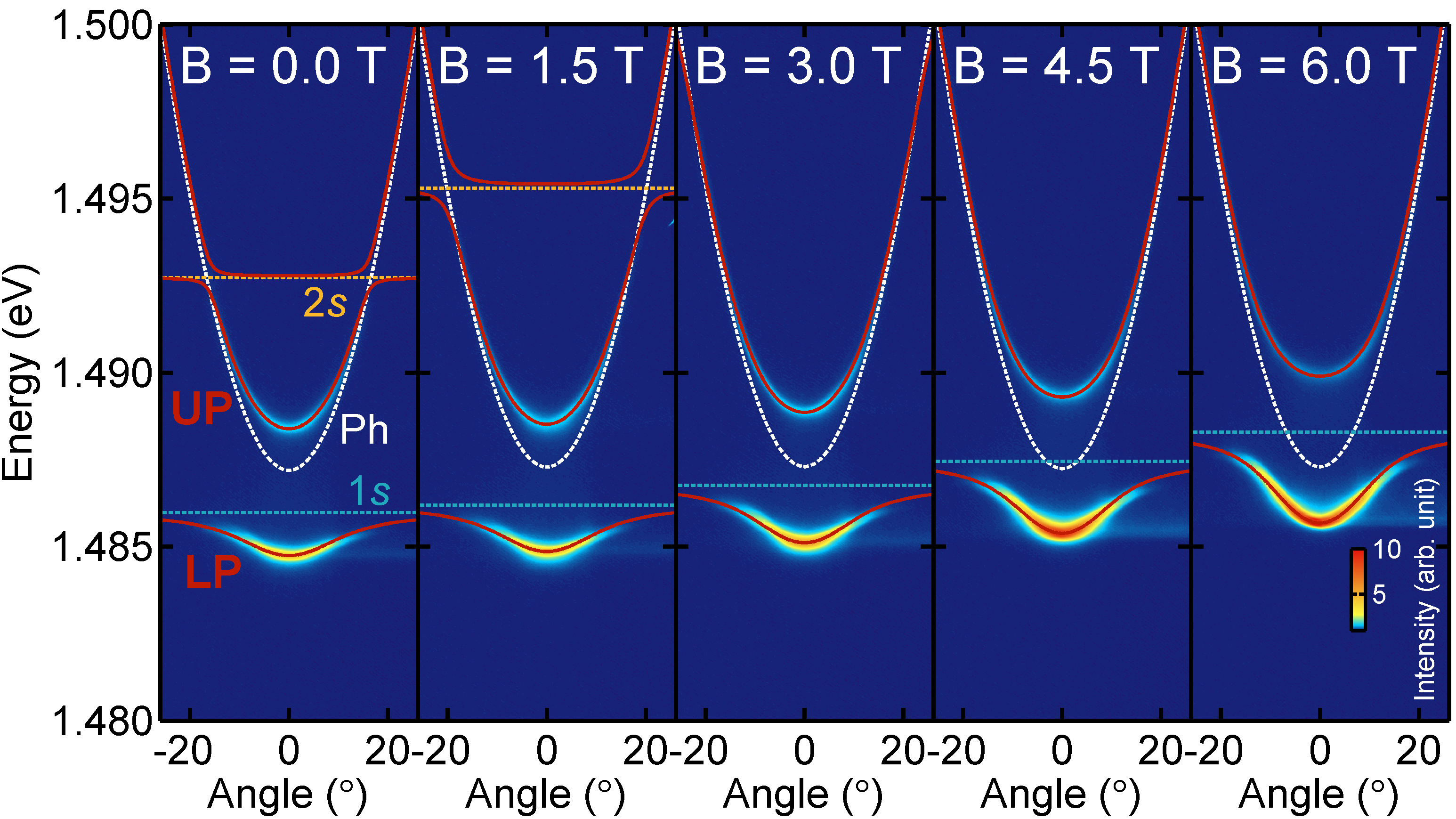}
	\caption{Angle-resolved photoluminescence spectra of exciton-polaritons in magnetic field. The solid red lines mark the polariton branches, the dotted lines indicate the calculated energies of uncoupled excitons: 1$s$ (blue) and 2$s$ (orange), and photon mode (white).}
	\label{f1}
\end{figure}
\begin{figure}
	\centering
	\includegraphics[width=8.7cm]{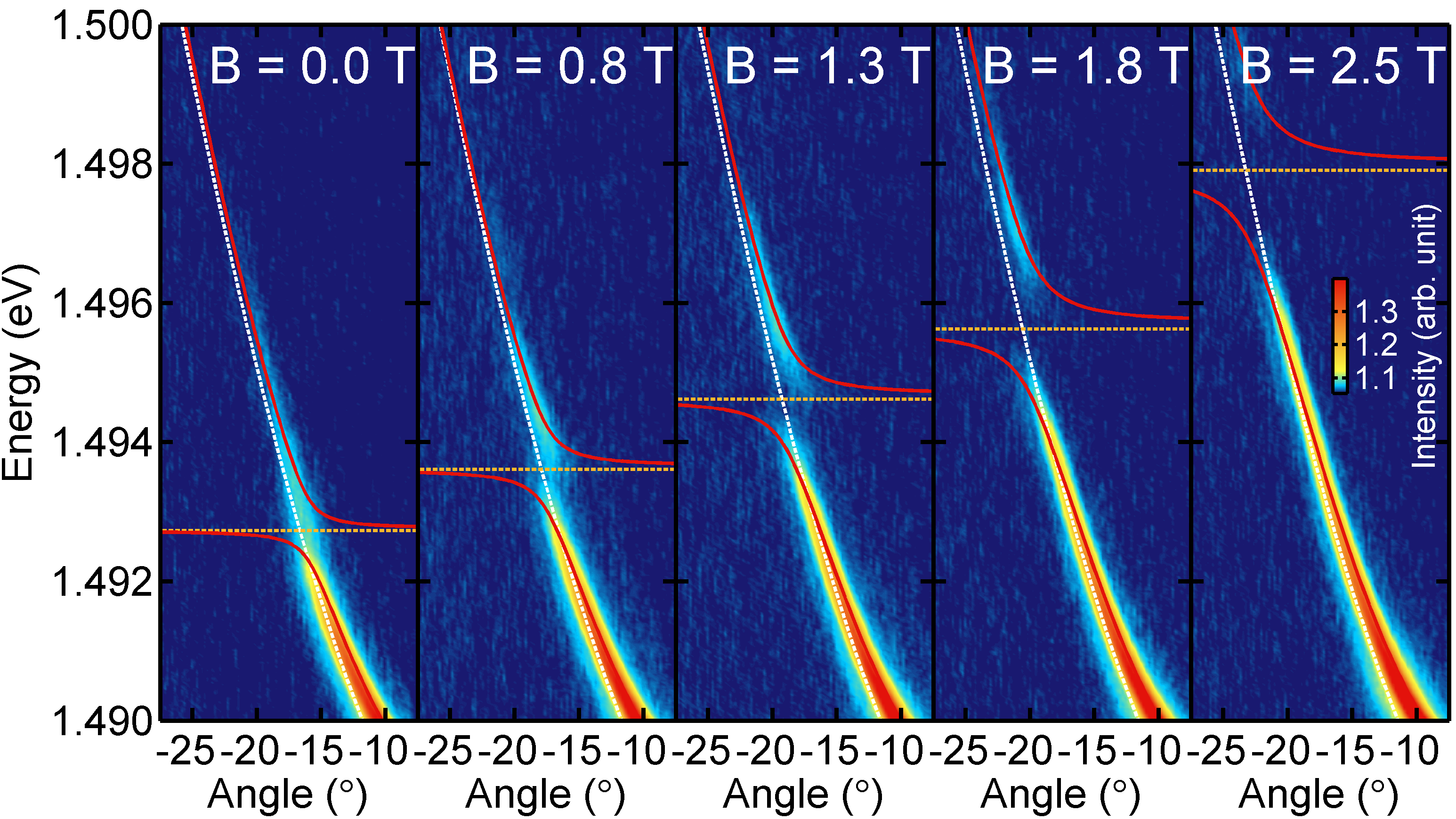}
	\caption{Angle-resolved photoluminescence spectra of exciton-polaritons in magnetic field in the low magnetic field regime for high emission angles, where the anti-crossing of photon mode with 2$s$ exciton is visible. The solid lines mark the position of the polariton branches, the dotted orange lines indicate the calculated energies of uncoupled 2$s$ exciton and photon modes. The blueshift in energy and the increase of the 2$s$ Rabi energy are observed.}
	\label{f2}
\end{figure}
\begin{figure}
	\centering
	\includegraphics[width=8.5cm]{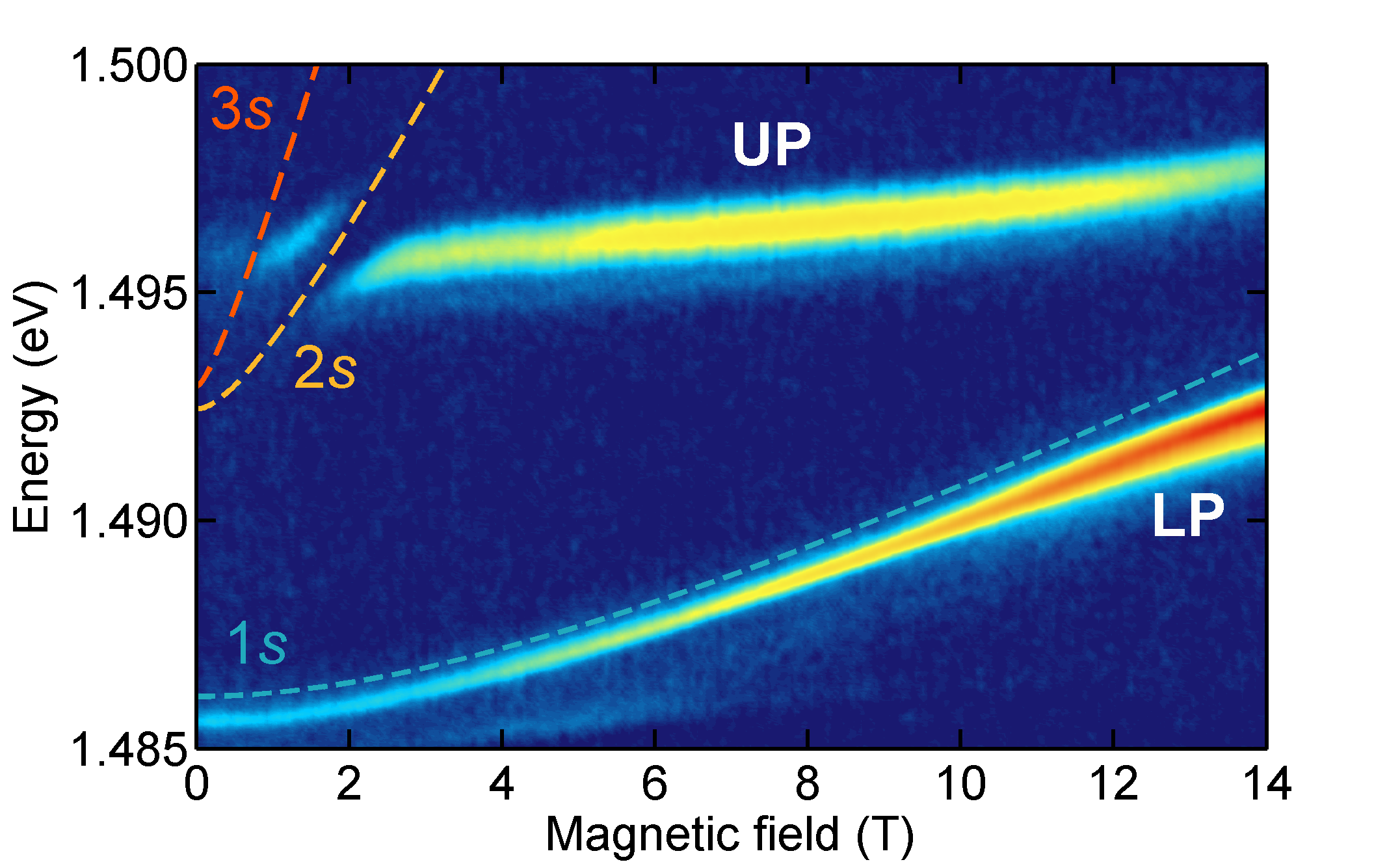}
	\caption{Photoluminescence of exciton-polariton at the emission angle of \SI{20}{\degree} over a wide range of magnetic fields. The anti-crossing observed on the UP is due to the coupling of the 2$s$ exciton state to cavity photons. The dashed lines mark the energy of 1$s$, 2$s$ and 3$s$ exciton states calculated with the use of the Pad\'e approximant method from Ref.~\cite{MacDonald}.}
	\label{f2a}
\end{figure}

\section{Exciton $2s$ state coupled to a cavity photon}
\label{sec:results}

The angle-resolved PL maps of exciton-polaritons in magnetic field up to 14~T are illustrated in Fig.~\ref{f1}. At low magnetic fields, below 3~T, we observe three polariton branches: the lower polariton (LP), the upper polariton (UP) and the UP anti-crossing with some additional resonance. The three branches are due to the coupling of a single cavity photon mode to two excitonic resonances in the quantum well: the 1$s$ ground state and the 2$s$ excited state. 
At higher magnetic fields (between 3~T and 14~T) only two polariton lines are visible. The magnetic field of approx. 3~T is the limit above which we do not observe anymore the 2$s$ excitonic state as it shifts towards higher energies and the coupling regime occurs for high emission angles, not detectable due to the limited numerical aperture of our experimental setup. 
Fig.~\ref{f2} illustrates the enlarged region of upper-polariton branch in the low magnetic field range and for emission angles where the 2$s$ exciton-polariton is visible. The increase of the 2$s$ exciton -- photon coupling strength is directly visible here as an increased separation between the polariton branches. The video illustrating the angle-resolved PL spectra in continuously changing magnetic field, in the region of 2$s$ exciton -- photon coupling, is provided in the Supplemental Information (SI). 

The 2$s$ exciton-polariton is also visible in Fig.~\ref{f2a}, which provides a view of the polariton emission at large emission angles as a function of magnetic field. The emission angle is kept constant and the polariton emission is traced in magnetic field. The additional anti-crossing observed at UP is due to a cavity photon coupling to 2$s$ exciton state. The signatures of a coupling to a higher 3$s$ exciton state is also visible, but the data do not allow for a clear resolution of this effect.

In order to describe these observations, we fit the experimental data to the theoretical model of exciton-photon coupling. When only two lines are visible (magnetic field stronger than 3~T), we use the standard polariton Hamiltonian. Excitons in the $1s$ state, confined inside the QW, couple to the photonic modes with the coupling strength given by the Rabi splitting, $\Omega_{1s}$. The coupling is described by a simple two-level Hamiltonian in the form
\begin{equation}
H=
\begin{pmatrix} 
E_{x}^{1s}(B) & \frac{\hbar}{2}\Omega_{1s} \\ 
\frac{\hbar}{2}\Omega_{1s} & E_{c}(k_{\parallel})
\end{pmatrix}
\label{eq0}
\end{equation}
where $E_{c}$ is the cavity photon energy and $E_{x}^{1s}$ is the exciton energy at zero momentum parallel to the cavity plane and at zero magnetic field.
In the case of our sample, the energy of the bare $1s$ excitonic resonance, $E_{x}^{1s}$, is at approx.~1.484~eV and the $1s$ exciton vacuum Rabi splitting is equal to approx.~$\Omega_{1s} = 3.5$~meV at zero magnetic field. 

Within this model we neglect the coupling to the light-hole excitons as they are not observed in this particular experiment. The heavy-light hole exciton energy separation in our QW is 16~meV and the strong coupling regime for light-hole exciton is observed only when photonic resonance is shifted towards higher energies (much more positive exciton - photon detuning).

In order to describe the coupling of two excitonic states, 1$s$ and 2$s$, to cavity photons, we extend the two-level coupled oscillator model given by~(\ref{eq0}) to a three-level model
\begin{equation}
H=
\begin{pmatrix} 
E_{x}^{1s}(B) & 0 & \frac{\hbar}{2}\Omega_{1s}\\ 
0 & E_{x}^{2s} & \frac{\hbar}{2}\Omega_{2s}\\
\frac{\hbar}{2}\Omega_{1s} & \frac{\hbar}{2}\Omega_{2s} & E_{c}(k_{\parallel})
\end{pmatrix}.
\label{eq1}
\end{equation}
where $E_{x}^{2s}$ is the 2$s$ exciton energy at zero momentum and at zero magnetic field, and the $\Omega_{2s}$ is the 2$s$ exciton-photon Rabi energy.

The experimental data were fitted using the three-level coupled oscillator model up to 3.0~T, Eq.~(\ref{eq1}) and two-level model, Eq.~(\ref{eq0}) from 3.0~T up to 14~T, which allows us to determine the LP and UP dispersion shapes together with the bare excitonic and photonic energies. The polariton modes and the bare resonances are plotted directly in Figs.~\ref{f1} and~\ref{f2}. The 1$s$ and 2$s$ exciton energy shift is shown in Fig.~\ref{f3} and the Rabi energies, $\Omega_{1s}$ and $\Omega_{2s}$, in Fig.~\ref{f4}. We note that the small jump of $\Omega_{1s}$ in Fig.~\ref{f4} at magnetic field 3~T is due to the change of the method of fitting from the two- to three-oscillator model.

\section{Theoretical model}

\begin{figure}
	\centering
	\includegraphics[width=7cm]{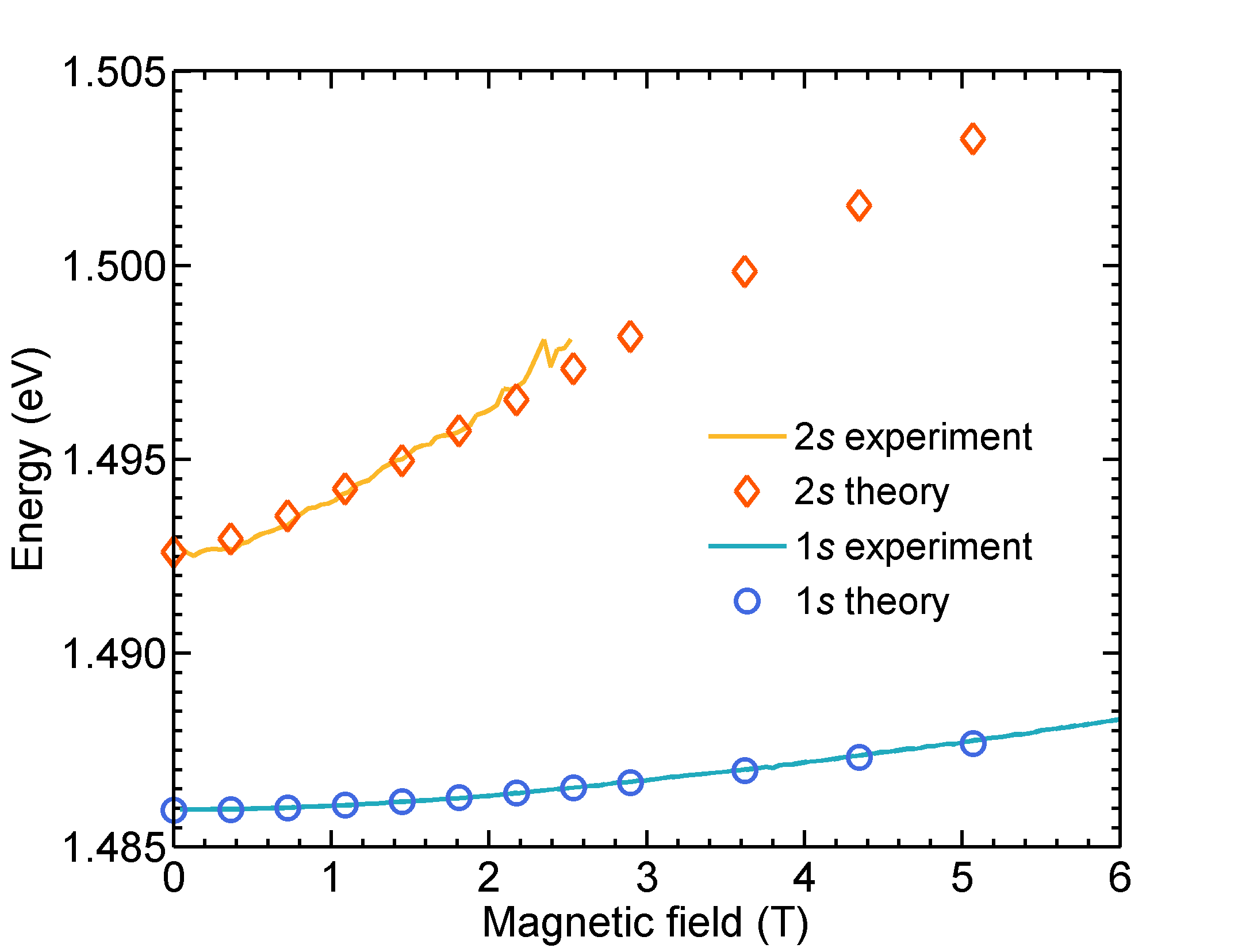}
	\caption{The energy levels of the 1$s$ and 2$s$ excitons in magnetic field. Solid lines represent the experimental data obtained from the fits of the two- and three-oscillator models, and points illustrate the energies of the 1$s$ and 2$s$ exciton states in the two-dimensional hydrogen atom model, calculated numerically using the imaginary time evolution method.}
	\label{f3}
\end{figure}

\begin{figure}
	\centering
	\includegraphics[width=7cm]{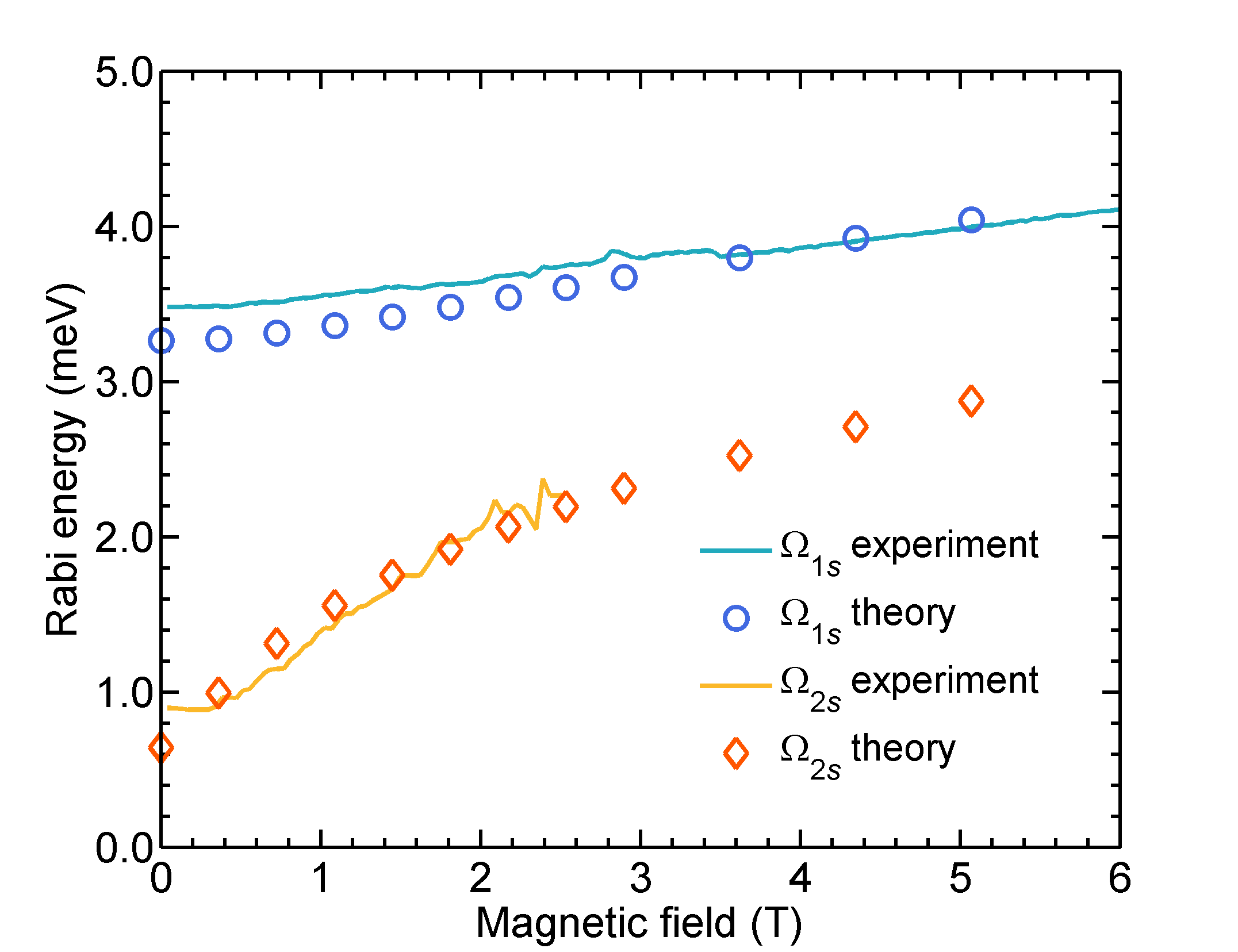}
	\caption{Increase of the Rabi energy in magnetic field for 1$s$ and 2$s$ excitons coupled to cavity photons. The lines and points have the same significance as in Fig.~\ref{f3}.}
	\label{f4}
\end{figure}

To describe the $1s$ and $2s$ QW exciton states and their coupling to light, we apply the model of a two-dimensional hydrogen-like atom described in Ref.~\cite{MacDonald}. We assume that the electron and the hole are tightly bound to the thin quantum well plane, which allows us to disregard the effects of transverse excitations. Due to the spatial invariance of the model in the plane of the QW,  the wave-function of the exciton can be further separated into the part corresponding to the center of mass motion, and the one corresponding to the relative electron-hole motion. The contribution of the center of mass motion to the energy of the exciton can be safely neglected, thanks to the large mass of the exciton as compared to the effective cavity photon mass. By applying these approximations we obtain a model of hydrogen-like atom described by the effective Hamiltonian
\begin{equation} \label{H_MD}
H = - \nabla^2-i\gamma \frac{m^*}{\eta}\frac{\partial}{\partial \phi} - \frac{2}{r} + \frac{\gamma^2r^2}{4},
\end{equation}
where $r$ and $\phi$ describe the relative positions of the electron and the hole, $m^*=1/(m_e^{-1} + m_h^{-1})$, $\eta^{-1}=m_e^{-1} - m_h^{-1}$, and $\gamma=\mu_B^* H /R_0^*$. Energy and spatial coordinates are expressed in units of the effective Rydberg constant $R_0^*=m^*e^4/2\epsilon^2 \hbar^2$ and exciton Bohr radius $a_0^*=\epsilon \hbar^2/m^*e^2$, respectively~\cite{MacDonald}. Note the slight difference in the second term of the Hamiltonian with respect to Ref.~\cite{MacDonald}, which comes from the fact that excitons are not simple atomic-like structures with a light electron and an heavy nucleus. Both electron and hole motions are important and both contribute to the interaction with magnetic field. This term, however, is not relevant to the $s$ states of excitons with zero angular momentum, discussed here.

The above model can be solved analytically in the limits of low magnetic field and high magnetic fields. In the latter the eigenstates become analogous to electronic Landau levels~\cite{MacDonald}. In the case of current experiment, however, neither of these limits is appropriate to describe the excitonic states. We resorted to numerical calculations of the eigenstates of the Hamiltonian~(\ref{H_MD}) using the projective imaginary time method. For each value of the magnetic field, the ground $1s$ exciton state was found numerically by evolution of the wave-function by imaginary time Schr\"odinger equation corresponding to Eq.~(\ref{H_MD}). The excited $2s$ state was obtained using the same method, in which additionally the state was projected in each simulation step to a subspace orthogonal to the previously found $1s$ state. This method is characterized by a good convergence and a relatively fast computation time.

The results of calculations are shown together with experimental data in Figs.~\ref{f3} and~\ref{f4}. The overall energy shift of the two lines, related to the energy bandgap, is treated as a free parameter. Although the agreement appears to be good in the whole range of magnetic fields, we point out that the $1s$ polariton was shifted with respect to the $2s$ polariton line upwards by approximately 5.9~meV. We explain this discrepancy between theory and experiment by the effect of transverse exciton excitations in a QW. We note that the above 2D model does not take into account the full three-dimensional structure of the exciton, neglecting the transverse degree of freedom. This assumption is not completely correct in the case of a relatively shallow and thin 8~nm-thick In$_{0.04}$Ga$_{0.96}$As QW, which corresponds to only about 5~meV confining potential for heavy holes and 60~meV for electrons. These values are comparable to the binding energy determined in the two-dimensional model, of the order of 15~meV and 2~meV for $1s$ and $2s$ states, respectively, which means that the transverse degree of freedom plays a significant role especially for the $1s$ exciton. Indeed, for a very shallow confining potential we expect the transition to a three-dimensional atom model. It is therefore interesting that the two-dimensional model can still describe very well the change of the exciton energy in function of the magnetic field.

The strength of the exciton-photon interaction is proportional to the probability of absorption of a photon and simultaneous creation of an electron-hole pair, or a reverse process of annihilation of an electron and a hole with accompanying creation of a photon. This probability is proportional, in the dipole approximation, to the amplitude of the relative electron-hole motion wave-function at the point of zero separation~\cite{Elliot}, $\Omega_i\sim |\psi^{e-h}_i(0)|$. These values were recovered from the eigenfunctions obtained by the imaginary time method and plotted against experimental data in Fig.~\ref{f4}. Also in this case, the agreement between theory and experiment is good (we did not correct for the transverse degree of freedom in the case of Rabi splitting).
The above dependence of Rabi energy on magnetic field shows very fast increase of $2s$ exciton-photon coupling, as compared to the $1s$ exciton, although the latter remains more strongly coupled for the whole range of magnetic fields where the $2s$ state could be resolved. 

The surprisingly strong increase of Rabi energy of the $2s$ exciton-polariton can be explained qualitatively in the framework of the perturbation theory applied in the limit of low magnetic fields. The Rabi energy of an $s$ state with quantum numbers $n$, $l=0$ is calculated as 
\begin{equation} \label{perturbation}
\Omega_{n,0}(B)\sim \left| \psi_{n,0}(0) + \frac{\gamma(B)^2}{4} \sum_{n'\neq n}\frac{\langle n', 0 | r^2 | n, 0 \rangle}{E_{n,0} - E_{n',0}} \psi_{n',0}(0) \right|_{B=0}
\end{equation}
where we have neglected $l\neq 0$ terms which do not contribute to the sum. The calculation taking into account up to four excited $s$ states gives $\Omega_{1s}\sim 1.60 + 0.01 \gamma^2$ and $\Omega_{2s}\sim 0.31 + 1.73 \gamma^2$. While the ratio of coupling strengths at zero magnetic field $\Omega_{1s}:\Omega_{2s} \approx 5:1$ is well reproduced (see Fig.~\ref{f4}), the ratio at higher magnetic fields is clearly overestimated. This demonstrates the limitation of the perturbation method, which cannot provide quantitative predictions even at moderate but nonzero magnetic fields, as opposed to the imaginary time numerical method. We also note that we found that the Pad\'e approximation method described in Ref.~\cite{MacDonald} failed to provide reasonable predictions in the intermediate magnetic field range.

Nevertheless, the perturbation method qualitatively predicts correctly the behavior of Rabi energy, which increases with magnetic field much faster for $2s$ than for the $1s$ exciton state. This can be simply explained by examining the terms in the formula~(\ref{perturbation}). The $1s$ state is strongly energetically separated from excited states according to the formula $E_n\sim(n-1/2)^{-2}$ valid for the 2D hydrogen atom, which gives the ratio 1:9 between the energy of the $1s$ and $2s$ states. This leads to large values of denominators in the sum, and weak perturbation of the wave-function of the $1s$ state. On the other hand, the $2s$ state is relatively close to the higher excited states ($3s$ and higher), which results in its contribution and strong modification of the wave-function in magnetic field. This is due to the fragility of the relatively weakly bound excited states as compared to the ground state. We note that this effect is especially pronounced in the case of 2D excitons in a quantum well, in comparison to the three-dimensional hydrogen atom for which $E_n\sim n^{-2}$, and the ratio of energies of the ground and excited state is only 1:4.

\section{Summary}

We have demonstrated the existence of a 2$s$ excited state of cavity exciton-polariton. The cavity photon can couple simultaneously to the ground (1$s$) and the excited state (2$s$) of the quantum well exciton. The strength of the coupling of the 2$s$ exciton is significantly increasing with the magnetic field, which is explained by a model of a two-dimensional hydrogen atom. 

Our observations play an important role when considering more complex exciton-polariton experiments performed on high quality GaAs-based cavities as for example the creation and manipulation of non-equilibrium polariton condensates~\cite{1,2,3,4}. The spontaneous creation of the excited cavity polariton state influences the polariton ground state and this additional resonance should be considered as an effective bright state channel in the creation and recombination process. This is particularly important in magnetic field, where the coupling strength to this excited state significantly increases.

\section{Acknowledgements}

This work was supported by the National Science Centre, Poland, under Projects No. 2011/01/D/ST7/04088, No. 2011/01/D/ST3/00482, No. 2015/17/B/ST3/02273 and No. 2015/18/E/ST3/00558. MK and RM acknowledge the support from Ministry of Higher Education budget for education as a research project "Diamentowy Grant": 0010/DIA/2016/45 and 0109/DIA/2015/44. Part of the experimental work was possible thanks to the Physics facing challenges of XXI century grant at the Warsaw University, Poland. MRM and MP acknowledge the support from ERC Advanced Research Grant MOMB No. 320590.



\end{document}